\documentclass[11pt]{article}
\usepackage{natbib,graphicx,setspace,amsmath,amssymb,subfigure}
\newtheorem{theorem}{Theorem}
\newtheorem{lemma}{Lemma}
\newtheorem{remark}{Remark}

\newtheorem{proposition}{Proposition}
\begin{document}

\title{Shrinkage Estimation and Selection for Multiple Functional Regression}         % Enter your title between curly braces
\author{Heng Lian\\Division of Mathematical Sciences\\School of Physical and Mathematical Sciences\\Nanyang Technological University\\Singapore, 637371\\E-mail: henglian@ntu.edu.sg}        % Enter your name between curly braces
\date{}          % Enter your date or \today between curly braces
\maketitle

\section*{Abstract} Functional linear regression is a useful extension of simple linear regression and has been investigated by many researchers. However, functional variable selection problems when multiple functional observations exist, which is the counterpart in the functional context of multiple linear regression, is seldom studied. Here we propose a method using group smoothly clipped absolute deviation penalty (gSCAD) which can perform regression estimation and variable selection simultaneously. We show the method can identify the true model consistently and discuss construction of pointwise confidence interval for the estimated functional coefficients. Our methodology and theory is verified by simulation studies as well as an application to spectrometrics data.\\
\textit{Keywords:} Estimation consistency, functional linear regression, group SCAD, principal component analysis, selection consistency.

\section{Introduction}       % Enter section title between curly braces
In several applications, functional data appear as the basic unit of observations. Classical regression models may be inadequate for such cases because of the high correlations for the discretized data. %As a natural extension of the multivariate data analysis, functional data analysis provides valuable insights into these problems. 
Compared with the discrete multivariate analysis, functional analysis takes into account the smoothness of the high dimensional covariates, and often suggests new approaches to the problems that have not been discovered before. %Even fornonfunctional data, the functional approach can often offer new perspectives on the old problem. 
Some recent developments in functional regression include \cite{yao05,cai06,crambes09,yuan09}.

The literature contains an impressive range of functional analysis tools for various problems.
%including exploratory functional principal component analysis, canonical correlation analysis, classification and regression. 
The more traditional
approach, carefully documented in the monograph \cite{ramsay05}, typically
starts by representing functional data by an expansion with respect to a certain basis, and
subsequent inferences are carried out on the coefficients. %The most commonly utilized basis include B-spline basis for nonperiodic data and Fourier basis for periodic data. 
Another line of work by the French school, taking a nonparametric point of view, extends the traditional
nonparametric techniques, most notably the kernel estimate, to the functional case \citep{ferraty06}. %Some theoretical results are also obtained as a generalization of the convergence properties of the classical kernel estimate. 
Besides these two major approaches, other methods such as putting functional regression in the reproducing kernel Hilbert space framework has been developed \citep{preda07,lian07}.

In this paper, we are concerned with an extension of simple functional linear regression model to the case where multiple functional observations are made on each unit. Formally, the model we consider is 
\begin{equation}\label{eqn:model}
Y_i=a+\sum_{j=1}^p \int_0^1 \beta_j(t)X_{ij}(t)\,dt+\epsilon_i, 1\le i \le n,
\end{equation}
where $X_{ij}$ are random functions, $a$ is the intercept, $\epsilon_i$ are random scalar errors and the functional coefficients $\beta_j, 1\le j\le p$ are the objects of interest in the model. 

Because functional coefficients are more complicated objects than scalar coefficients in classical multiple linear regression, it is generally desirable to identify those significant variables in predicting the responses, even if $p$ is small. For example, in a similar context, \cite{zhu10} investigated fluorescence
spectroscopy for cervical precancer diagnosis, where a Bayesian model is used to select from multiple fluorescence spectra for classification of subjects. 

In a non-Bayesian context, traditional methods for variable selection in classical linear models include constructing hypothesis tests or using information criteria. More recently, regularization methods have received much attention. For standard linear regression, Lasso \citep{tibshirani96} is
probably the most popular method that uses $L_1$ penalty to force some of the coefficients to be equal to zero. %Compared to traditional model selection method using information criteria, Lasso is continuous and thus more stable. More systematic theoretical studies on Lasso appeared later. [9] showed that Lasso is consistent for prediction, a property that was called persistency. 
Several subsequent works \citep{meinshausen06,zhaoyu06} have shown that Lasso is in general not consistent for model selection unless some nontrivial conditions on the covariates are satisfied. %Even when those conditions are satisfied, the efficiency of the estimator is compromised when one insists on variable selection consistency since the coefficients are over-shrinked. 
To address these shortcomings of Lasso, \cite{fan01} proposed the smoothly clipped absolute deviation (SCAD) penalty %which takes into account several desired properties of the estimator like continuity, asymptotic unbiasedness, etc. %They also show that the resulting estimator possesses the oracle property, i.e. it is consistent for variable selection and behaves the same as when the zero coefficients are known in advance. These results are extended to the case with a diverging number of covariates in [5]. In \cite{zou06}, the author 
and \cite{zou06} proposed the adaptive lasso in the fixed $p$ case using a weighted $L_1$ penalty with weights determined by an initial estimator, %The idea behind the adaptive lasso is to assign higher penalty for zero coefficients and lower penalty for larger coefficients. [12] studied the adaptive lasso with a diverging number of parameters and proposed using marginal regression as the initial estimator under partial orthogonality assumption. Also in the high dimensional case, [11] showed similar oracle properties for the estimator with L  penalty when 0 < ¦Ã < 1. 
and there are many other extensions of the regularization framework for variable selection \citep{yuan06,wangleng07,huangma08}.

In this article we use functional principle component analysis (PCA)-based estimation method \citep{cardot99,hall07} combined with group SCAD (this terminology seems to first appear in \cite{wang07} for varying coefficients variable selection) which represents a new application of the SCAD penalty.  %to multiple functional regression. 
Regularization method for variable selection in nonparametric settings has been developed in the context of smoothing spline ANOVA for nonparametric regression smoother \citep{lin06} and support vector machines \citep{zhang06}.  For varying-coefficient models,  we are aware of the work \citep{wang08} where the authors used basis expansion approach for estimation combined with group SCAD penalty on coefficients, and the work \citep{wang09} where group Lasso penalty is applied directly to smooth functions evaluated at sampled points.%, both methods effectively shrink irrelevant functional variables to achieve model selection consistency. 

The rest of the article is organized as follows. We describe the functional PCA and shrinkage estimation procedure in Section 2.1, and present estimation consistency and selection consistency results in Section 2.2. Then we discuss estimation and inference algorithms and tuning parameter selection in Section 2.3 and 2.4 respectively. In Section 3, we present some simulation experiments and illustrate the proposed method using the spectrometrics data example where the goal is to predict the percentage of fat content in the piece of meat based on spectra curves. All technical details are gathered in the Appendix.

\section{Methodology and Theoretical Properties}
\subsection{Estimation of Multiple Functional Regression}
Assume we have independent and identically distributed (i.i.d.) observations $(X_{11},\ldots,X_{ip},Y_i)$, $1\le i\le n$, where $X_{ij}$ is a square integrable random function on the interval $[0,1]$ with mean $\mu_j$. The response variables $Y_i$ are generated by model (\ref{eqn:model}) with i.i.d. errors $\epsilon_i$ of finite second moments. We also assume the errors are independent of the predictors. We will use $(X_1,\ldots,X_p,Y)$ to denote the generic random variables with distribution the same as $(X_{i1},\ldots,X_{ip},Y_i)$. Let $K_j(s,t)=Cov\{X_j(s),X_j(t)\}$, and by Mercer's theorem we have the spectral expansion 
\[K_j(s,t)=\sum_{k=1}^\infty \lambda_{jk}\phi_{jk}(s)\phi_{jk}(t),\]
where $\lambda_{j1}>\lambda_{j2}>\cdots>0$ are the eigenvalues of the linear operator associated with $K_j(s,t)$ with corresponding eigenfunctions $\phi_{jk}$. 
On the other hand, let $\hat{K}_j(s,t)=\frac{1}{n}\sum_{i=1}^n(X_{ij}(s)-\bar{X}_j(s))(X_{ij}(t)-\bar{X}_j(t))$  where $\bar{X}_j=\sum_i X_{ij}/n$, we have the empirical counterpart of the above expansion,
\[\hat{K}_j(s,t)=\sum_{k=1}^\infty \hat{\lambda}_{jk}\hat{\phi}_{jk}(s)\hat{\phi}_{jk}(t),\]
where $\hat{\lambda}_{j1}\ge\hat{\lambda}_{j2}\ge\cdots\ge 0$. To get rid of uncertainty of signs, we assume $\int \hat{\phi}_j\phi_j\ge 0$. For the empirical operator $\hat{K}$, at most $n$ eigenvalues are strictly positive. 

In general, different functional predictors are not independent of each other. The Karhunen-Lo\`{e}ve expansion of the random function $X_{ij}$ in terms of the orthonormal basis $\phi_{jk}(t)$ is given by
\begin{equation}\label{eqn:KL}
X_{ij}-\mu_j=\sum_{k=1}^\infty\xi_{ijk}\phi_{jk},
\end{equation}
where $\xi_{ijk}$ are principal component scores satisfying $E\xi_{ijk}=0, E\xi_{ijk}^2=\lambda_{jk}$ and $E\xi_{ijk}\xi_{ijk'}=0, k\neq k'$. Thus from (\ref{eqn:KL}) we have the covariance operator expansion
\[Cov\{X_{j_1}(s),X_{j_2}(t)\}=\sum_{k_1,k_2=1}^\infty \lambda^{j_1,j_2}_{k_1,k_2}\phi_{j_1k_1}(s)\phi_{j_2k_2}(t),\]
where $\lambda^{j_1,j_2}_{k_1,k_2}=E\xi_{j_1k_1}\xi_{j_2k_2}$ determines the dependency structure between different predictors. Note with our notation, when $j_1=j_2=j$, $\lambda^{j,j}_{k_1,k_2}=0$ if $k_1\neq k_2$ and $\lambda^{j,j}_{k_1,k_2}=\lambda_{jk}$  if $k_1=k_2=k$. An illustration of how this dependency could arise is given in the next subsection. 

The model (\ref{eqn:model}) can be equivalently written as
\begin{equation}\label{eqn:model2}
Y_i-\mu=\sum_{j=1}^p\int\beta_j(X_{ij}-\mu_j)+\epsilon_i,
\end{equation}
where $\mu=E[Y|X_1,\ldots,X_p]=a+\sum_j\int\beta_j\mu_j$. After $\beta_j$ is estimated by $\hat{\beta}_j$, say, the intercept $a$ can be easily estimated by $\hat{a}=\bar{Y}-\sum_j\int\hat{\beta}_j\bar{X}_j$, where $\bar{Y}=\sum_i Y_i/n$.

Now we consider the problem of estimating $\beta_j$. Using the orthonormal basis $\{\phi_{jk}\}$, (\ref{eqn:model2}) can be equivalently written as 
\[Y_i-\mu=\sum_{j=1}^p\sum_{k=1}^\infty\xi_{ijk}b_{jk}+\epsilon_i,\]
making use of the expansion $\beta_j=\sum_k b_{jk}\phi_{jk}$. This suggests the estimator
\[\{\hat{b}_{jk}\}=\arg\min \sum_{i=1}^n (Y_i-\bar{Y}-\sum_{j=1}^p\sum_{k=1}^K\hat{\xi}_{ijk}b_{jk})^2,\]
and then $\hat{\beta}_j=\sum_{k=1}^K\hat{b}_{jk}\hat{\phi}_{jk}$, where in the above displayed equation $\hat{\xi}_{ijk}=\int (X_{ij}-\bar{X}_j)\hat{\phi}_{jk}$ is the principal component score estimated from data. Here the truncation point $K$ is a smoothing parameter. To further select functional predictors simultaneously, we minimize the criterion function
\begin{equation}\label{eqn:J}
J(b)=\sum_{i=1}^n (Y_i-\bar{Y}-\sum_{j=1}^p\sum_{k=1}^K\hat{\xi}_{ijk}b_{jk})^2+n\sum_{j=1}^p p_\lambda(||b_j||),
\end{equation}
where $||b_j||$ is the $l_2$ norm of $b_j=(b_{j1},\ldots,b_{jK})^T$. Among many ways to specify the penalty function $p_\lambda$, we choose the SCAD penalty function of \cite{fan01}, which can be defined by
\[p'_\lambda(\theta)=\lambda\left\{I(\theta\le\lambda)+\frac{(a\lambda-\theta)_+}{(a-1)\lambda}I(\theta>\lambda)\right\},\;p_\lambda(0)=0,\]
for $a=3.7$ and $\theta>0$, where $I(\cdot)$ is the indicator function. The choice of $a=3.7$ is suggested by \cite{fan01} and adopted in almost all publications involving SCAD penalty. Other penalty functions such as the adaptive Lasso can also be used here and will lead to similar consistency results as below. 

\subsection{Consistency Properties}
Large sample properties of shrinkage estimation with SCAD penalty have been established in the literature \citep{fan01,fan04,wang08}. We show that in our context the estimation procedure can consistently estimate the functional coefficients as well as consistently identify the true model. However, extending these theoretical results to multiple functional regression is not trivial. Note that in criterion (\ref{eqn:J}) two types of approximations are involved, one is the truncation of $\beta_j$ to approximate the functional coefficients, the other is the unknown covariate $\xi_{ijk}$ estimated by $\hat{\xi}_{ijk}$. While the former approximation is typical in nonparametric problems such as \cite{wang08}, the latter is unique to the functional regression problem. It also resembles the measurement error model in form where the covariates are not observed directly \citep{liang09,carroll09}. 

We denote the true regression coefficients by $\beta=((\beta^{(1)})^T,(\beta^{(2)})^T)^T$ with $\beta^{(1)}=(\beta_1,\ldots,\beta_s)^T, s\le p$ containing all nonvanishing components of $\beta$ and $\beta_{s+1}=\cdots=\beta_p\equiv 0$. Let $\Lambda$ be the $pK\times pK$ matrix
\begin{equation}\label{eqn:Lambda}
\left(\begin{array}{ccc}
		\Lambda^{1,1}&\cdots&\Lambda^{1,p}\\
		\vdots&\vdots&\vdots\\
		\Lambda^{p,1}&\cdots&\Lambda^{p,p}\\
	\end{array}\right)\;,
\end{equation} 
where $\Lambda^{j_1,j_2}$ is the $K\times K$ matrix with entries $\lambda^{j_1,j_2}_{k_1,k_2}, 1\le k_1,k_2\le K$.
In our results, the following regularity conditions are needed:
\begin{enumerate}
\item[(c1)] $X_{ij}$ has finite fourth moment: $\int E(X_{ij}^4)<\infty$, and $E\epsilon_i^4<\infty$.
\item[(c2)] $\lambda_k-\lambda_{k+1}\ge C^{-1}k^{-\alpha-1}$, $b_{jk}\le Ck^{-\beta}$, $\alpha>1,\beta>\alpha+1/2$.
\item[(c3)] The smoothing parameter $K$ satisfies $n^{-1/(5\alpha+3)}K\rightarrow 0$, and $n^{-1/(3\alpha+2\beta+2)}K$ is bounded away from $0$. 
\item[(c4)] The parameter $\lambda$ satisfies $\lambda=o(K^{-\alpha})$ and $\sqrt{K^{3\alpha+3}/n}=o(\lambda)$.
\item[(c5)] The minimum eigenvalue of $\Lambda$, denoted by $\rho_{\min}(\Lambda)$, is of order $\Omega(K^{-\alpha})$ where $a_n=\Omega(b_n)$ means $b_n=O(a_n)$. 
\end{enumerate}
\begin{remark} Because $\beta>\alpha+1/2$, there exists $K$ satisfying (c3). Also, we have $\sqrt{K^{3\alpha+3}/n}=o(K^{-\alpha})$ when $K$ satisfies (c3) and thus there exists $\lambda$ satisfying (c4).
\end{remark}
Based on the assumptions listed above, we can establish the following result.
\begin{theorem} \label{thm:main}Assume (c1)-(c5), we have

(a)(Estimation consistency) $||\hat{\beta}_j-\beta_j||=o_p(1), 1\le j\le p$.

(b)(Selection consistency) $\hat{\beta}_{s+1}=\ldots=\hat{\beta}_p\equiv 0$ with probability converging to $1$.

\end{theorem}
Note that the study of optimal convergence rates for multiple functional regression problem is more complicated and we do not attempt it here. 

\textbf{An illustration.} Let $p=2$. Suppose the eigenvalues of $K_1$ and $K_2$ satisfy $\lambda_{jk}=Ck^{-\alpha}, j=1,2.$ If $X_1$ and $X_2$ are independent, then the matrix $\Lambda$ defined in (\ref{eqn:Lambda}) is diagonal and its minimum eigenvalue is obviously of order $\Omega(K^{-\alpha})$. In general, $\Lambda$ can be written as a block matrix
\[\Lambda=\left(\begin{array}{cc}
		E & F\\
		F^T & G
		\end{array}\right)\;,
\]
where $E$ and $G$ are $K\times K$ diagonal matrices containing the eigenvalues of $K_1$ and $K_2$ respectively. It is easy to see that the minimum eigenvalue of $\Lambda$ is no bigger than $CK^{-\alpha}$, since $\Lambda$ is similar to 
\[ \tilde{\Lambda}=\left(\begin{array}{cc}
		E & 0\\
		0 & G-F^TE^{-1}F
		\end{array}\right)
\] (that is $\Lambda=P^{-1}\tilde{\Lambda}P$, for some invertible matrix $P$), and obviously the eigenvalues of $G-F^TE^{-1}F$ are dominated by those of $G$. In assumption (c5), we assume that the minimum eigenvalue of $\Lambda$ is still of order $K^{-\alpha}$ as in the independent case. This assumption thus can be thought of as a constraint on the dependence of different predictors. However, we show in the following setup this assumption is quite natural. Suppose the random functions $X_1$ and $X_2$ are specified by
\begin{equation}\label{eqn:trans}
X_1=\sum_{j=1}^l a_{1j}W_j, X_2=\sum_{j=1}^l a_{2j}W_j,
\end{equation}
where $W_j, 1\le j\le l$ are independent mean zero random functions with Karhunen-Lo\`{e}ve expansion give by $W_j=\sum_k\omega_{jk}\phi_k$ (note that we assume the eigenfunctions are common to all $W_j$) with $E\omega_{jk}^2=\kappa_{jk}>0$. The following proposition gives a sufficient condition under which $\rho_{\min}(\Lambda)=\Omega(K^{-\alpha})$. 

\begin{proposition}\label{prop:1} Suppose $ck^{-\alpha}\le \kappa_{jk}\le Ck^{-\alpha}, j=1,\ldots, l$, for some constants $C\ge c>0$. If $\{a_{1j}\}, \{a_{2j}\}$ are two fixed sequences and one is not a scalar multiple of the other (that is, for any constant $\gamma$, we cannot have $a_{1j}=\gamma a_{2j}$ or $a_{2j}=\gamma a_{1j}$ for all $j$), then $\rho_{\min}(\Lambda)=\Omega(K^{-\alpha})$. 
\end{proposition}

\subsection{Computation and Inferences}
One can express the criterion function $J(b)$ in vector and matrix form. Let 
\[\hat{Z}_j=\left(\begin{array}{ccc}
		\hat{\xi}_{1j1}&\ldots&\hat{\xi}_{1jK}\\
		\vdots&\vdots&\vdots\\
		\hat{\xi}_{ij1}&\ldots&\hat{\xi}_{ijK}\\
		\vdots&\vdots&\vdots\\
		\hat{\xi}_{nj1}&\ldots&\hat{\xi}_{njK}\\
		\end{array}\right),\]
$\hat{Z}=(\hat{Z}_1,\ldots,\hat{Z}_p)$, $b=(b_{11},\ldots,b_{1K},b_{21},\ldots,b_{pK})^T$, $\mathbf{Y}=(Y_1,\ldots,Y_n)^T$, the criterion (\ref{eqn:J}) is written as
\begin{equation}\label{eqn:J2}
J(b)=\sum_{i=1}^n(\mathbf{Y}-\bar{Y}\mathbf{1}-Zb)^T(\mathbf{Y}-\bar{Y}\mathbf{1}-Zb)+n\sum_{j=1}^p p_\lambda(||b_j||),
\end{equation}
where $\mathbf{1}$ is the n-dimensional vector with all components ones.

We use the local quadratic approximation idea \citep{fan01} to optimize the criterion. Specifically, if $\hat{b}^{(m)}$ is the estimate obtained in the $m$-th iteration, then the criterion (\ref{eqn:J2}) can be locally approximated by 
\begin{equation}\label{eqn:lqa}
\sum_{i=1}^n(\mathbf{Y}-\bar{Y}\mathbf{1}-\hat{Z}b)^T(\mathbf{Y}-\bar{Y}\mathbf{1}-\hat{Z}b)+\frac{n}{2}\sum_j b_j^T R(\hat{b}^{(m)})b_j,
\end{equation}
where $R(\hat{b}^{(m)})=diag\{(p'_\lambda(||\hat{b}_1^{(m)}||)/||\hat{b}_1^{(m)}||)I_K,\ldots,(p'_\lambda(||\hat{b}_p^{(m)}||)/||\hat{b}_j^{(m)}||)I_K\}$ and $I_K$ is the $K\times K$ identity matrix. The minimizer of (\ref{eqn:lqa}) is then given by 
\[\hat{b}^{(m+1)}=(\hat{Z}^T\hat{Z}+nR(\hat{b}^{(m)}))^{-1}\hat{Z}^T(\mathbf{Y}-\bar{Y}\mathbf{1}).\]
We iterate these steps until convergence and obtain the final estimate $\hat{b}$. During the iterations, depending on the choice of $\lambda$, if some $||b_j||$ is smaller than a threshold ($10^{-5}$ in our implementation), then we set $b_j=0$ and ignore the corresponding predictor in future iterations.  

Now we consider the construction of pointwise confidence intervals for $\beta_j$. Following \cite{fan01}, the sandwich formula can be used as an estimator for the variance
of the nonzero components of $\hat{b}$, denoted by $\hat{b}^{(1)}$ henceforth. The estimator of asymptotic variance is given by
\[\widehat{Cov}(\hat{b}^{(1)})=((\hat{Z}^{(1)})^T\hat{Z}^{(1)}+nR^{(1)})^{-1}(\hat{Z}^{(1)})^T\widehat{Cov}(Y)\hat{Z}^{(1)}((\hat{Z}^{(1)})^T\hat{Z}^{(1)}+nR^{(1)})^{-1},\]
where $\hat{Z}^{(1)}$ denotes the selected columns of $\hat{Z}$ corresponding to nonvanishing $||b_j||$, $R^{(1)}$ denotes the selected rows and columns of $R(\hat{b})$ in a similar way, and $\widehat{Cov}(Y)$ is the $n\times n$ diagonal matrix with estimated squared residuals on the diagonal. The diagonal blocks of $\widehat{Cov}(\hat{b}^{(1)})$ gives the asymptotic variance for nonvanishing $\hat{b}_j$.

Since $\hat{\beta}_j(t)=\hat{b}_j^T\hat{\phi}_j(t)$, $\hat{\phi}_j(t)=(\hat{\phi}_{j1}(t),\ldots,\hat{\phi}_{jK}(t))^T$. We have the natural estimator for the asymptotic variance of $\beta_j(t)$:
\[\widehat{Cov}(\beta_j(t))=\hat{\phi}_j(t)^T\widehat{Cov}(\hat{b}_j)\hat{\phi}_j(t).\]
Note that here we ignored the uncertainty of $\hat{\phi}_j$ which is also estimated from observations. However, we think this might be a reasonable first approximation because one might argue that ${\phi}_{jk}$ is easier to estimate than $\beta_j$ since $\phi_{jk}$ needs to be estimated first and its accuracy affect the subsequent estimations. Of course this is just a heuristic argument and we later use our simulation experiments to illustrate the performance the asymptotic variance formula. Estimates of the asymptotic variance can be used to construct pointwise confidence intervals for $\beta_j(t)$ for nonzero components of the functional coefficients. Strictly speaking, the constructed intervals will only be for the truncated $\beta_j(t)$ at cutoff $K$ in the expansion. Thus the constructed intervals will have lower than targeted coverage rate: on the one hand the variability in $\hat{\phi}_{jk}$ is ignored, on the other hand the interval is only for truncated functional coefficients. The bias caused will be seen from our numerical results.

\subsection{Tuning Parameter Selection}
For implementation of our method, we need to choose two smoothing parameters, the truncation point $K$ and the regularization parameter $\lambda$ for group SCAD penalty. In fact, for different predictors we might choose a different truncation point, which nevertheless would lead to a significant increase in computational burden, and thus we constrain the truncation point to be the same for all predictors. Besides, one can argue that the sensitivity of the estimator to the choice of $K$ is reduced by the extra smoothing parameter $\lambda$ in regularized estimation. 

We use generalized cross-validation (GCV) to select both $K$ and $\lambda$. GCV can be thought of as a short-cut for leave-one-out cross-validation, and also comes with advantageous properties \citep{wahba90}. The GCV criterion is defined by
\[GCV(K,\lambda)=\frac{1}{n}\frac{||\mathbf{Y}-\bar{Y}\mathbf{1}-\hat{\mathbf{Y}}||^2}{\left(1-tr(H(K,\lambda))/n\right)^2},\]
where $\hat{\mathbf{Y}}=H(K,\lambda)(\mathbf{Y}-\bar{Y}\mathbf{1})$ is the fitted response values and $H(K,\lambda)=\hat{Z}(\hat{Z}^T\hat{Z}+nR(\hat{b}))^{-1}\hat{Z}^T$ is the hat matrix.
\section{Numerical Experiments}
\subsection{Simulation Examples}
We perform a Monte Carlo experiment to investigate the finite sample performance of the estimation method, using GCV to select the two tuning parameters. Our example follow the illustration presented after Theorem \ref{thm:main}. The simulated data is generated from model (\ref{eqn:model}) with $p=4$ functional predictors, $a=0$ and the errors $\epsilon$ distributed as $N(0,\sigma^2)$. For $1\le j\le 4$ independently, we take $W_j=\sum_{k=1}^{50}\xi_{jk}\phi_k$ where $\xi_{jk}\sim N(0,k^{-2})$, $\phi_1\equiv 1$ and $\phi_{k+1}=\sqrt{2}\cos(k\pi t)$ for $k\ge 1$. Then the functional predictors are defined through the linear transformations
\begin{eqnarray*}
X_1&=&W_1+\rho(W_2+W_3),\\
X_2&=&W_2+\rho(W_1+W_3),\\
X_3&=&W_3+\rho(W_1+W_2),\\
X_4&=&W_4.
\end{eqnarray*}
Note the scalar $\rho$ controls the strength of dependence between different predictors, with $\rho=0$ resulting in independent predictors.  For $\beta_1$ and $\beta_2$, in terms of expansion based on $\{\phi_k\}$, we take $b_1=(-2,1,-2,1)^T$, $b_2=(1,-1,0.5,-0.5)^T$ and set $\beta_3=\beta_4=0$. We fix $n=100$ for all our simulations and set $\rho=0, 0.2 $ or $ 0.5 $, and $\sigma^2=0.1 $ or $0.3$. All integrations required in the generation of the data and the estimation procedure are performed using a Riemannian sum approximation with an equally spaced grid containing $500$ points on $[0,1]$.

The simulation results are summarized in Table \ref{tab:sim} based on $500$ runs in each scenario, where we report the mean squared errors $||\hat{\beta}-\beta||^2$ using our regularized multiple functional regression model (MSE), oracle mean squared errors where the true zero coefficients are known and no shrinkage is applied (OMSE), average number of correctly identified nonvanishing coefficients (TP), average number of incorrectly identified nonvanishing coefficients (FP),  empirical coverage probability of pointwise 95\% confidence interval for ${\beta}_1$ (95\% Cov.Prob.1) and empirical coverage probability of pointwise 95\% confidence interval for ${\beta}_2$ (95\% Cov.Prob.2).   For each scenario, the empirical coverage probabilities reported are the averages over the grid $(0.1,0.2,\ldots,0.9)$ for $\beta_1$ and $\beta_2$ whenever they are estimated as nonzero coefficients.

\begin{figure}
\centerline{\subfigure[]{\includegraphics[width=2.5in]{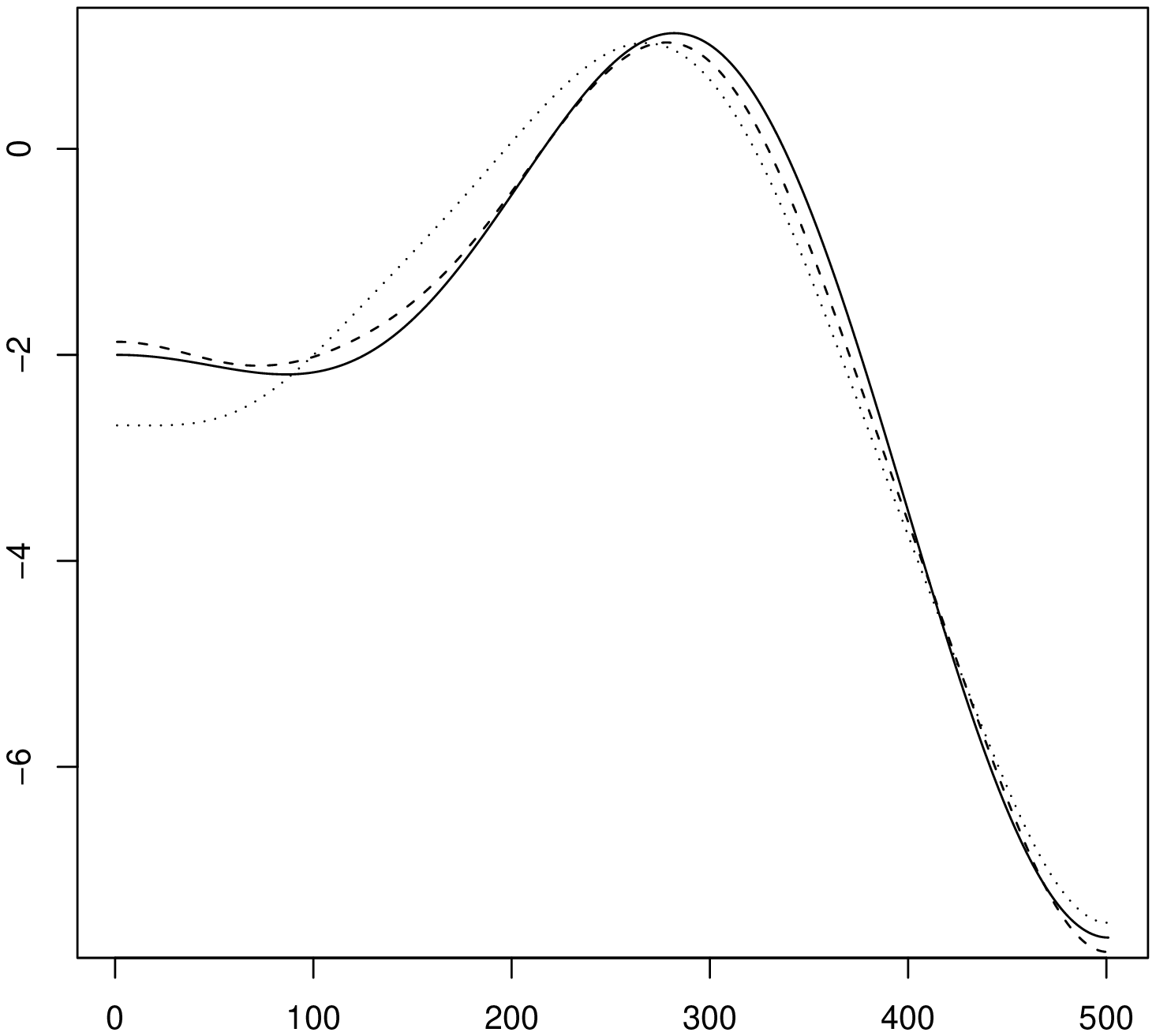}
}
\hfil
\subfigure[]{\includegraphics[width=2.5in]{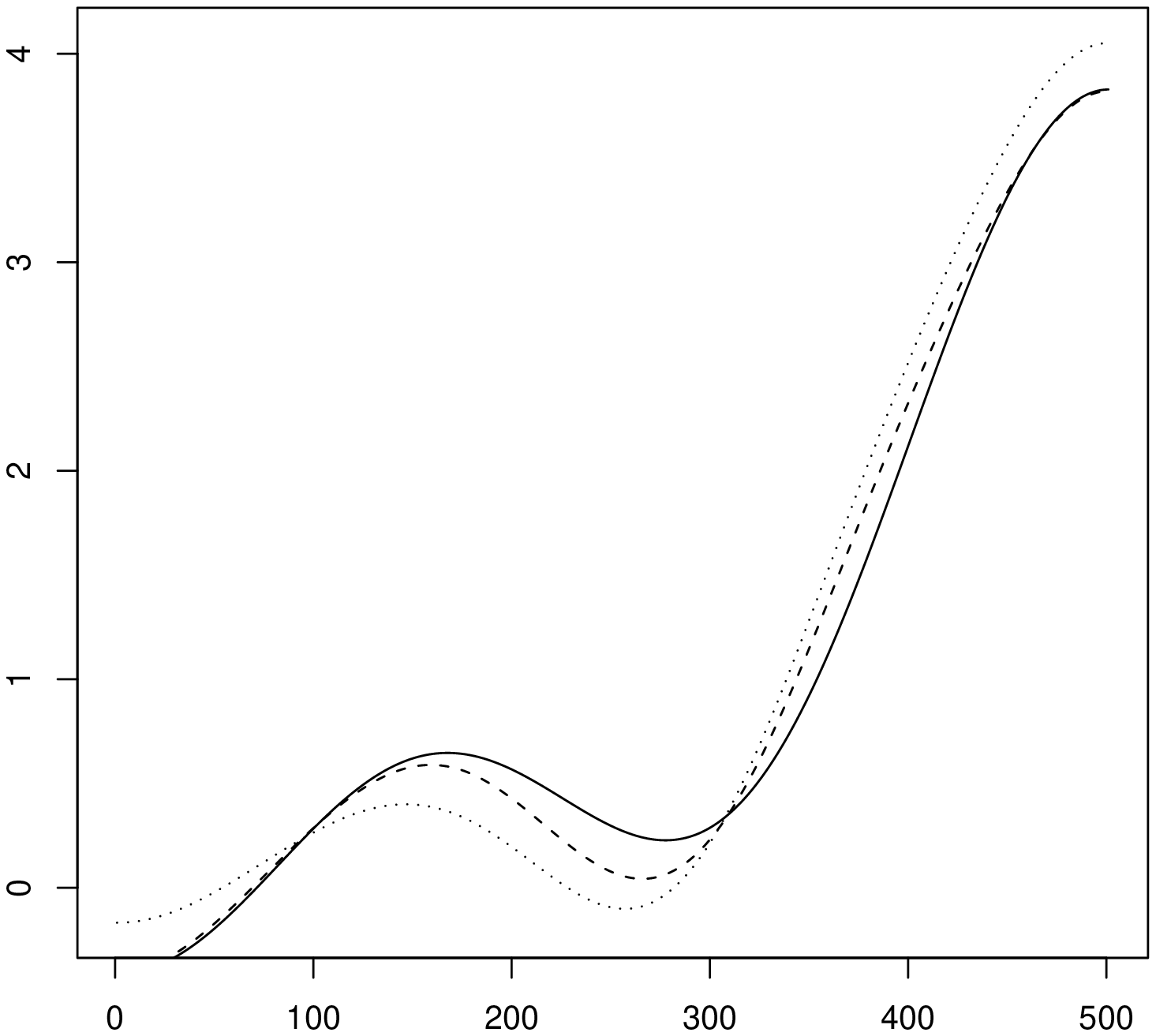}
}}
\caption{(a) The true coefficient $\beta_1$ (solid line) with its estimates when $\sigma=0.1$ (dashed line) and $\sigma=0.3$ (dotted line). (b) $\beta_2$ (solid line) with its estimates when $\sigma=0.1$ (dashed line) and $\sigma=0.3$ (dotted line). Here we set $\rho=0.2$.\label{fig:b}   }
\end{figure}

As one can see from Table \ref{tab:sim}, the noise level clearly has a significant effect on the estimation errors as well as the average number of truly relevant predictors detected. However, the number of false positives remains at a low level even for larger noise variance. Compared to noise level, the correlation between different predictors seems to have milder effects, with estimation error increasing with correlation strength. The result also shows that confidence intervals constructed based on the sandwich formula for the asymptotic variance work surprising well, with only a small downward bias in our simulations. As an illustration, the true functions $\beta_1$, $\beta_2$ as well as their estimates when $\rho=0.2$ and $\sigma=0.1$ or $0.3$ are plotted in Figure \ref{fig:b}.

%\begin{table}
%  \caption{Simulation results for penalized multiple functional regression. \label{tab:sim}}
%\bigskip
%\centering{\begin{tabular}{ccccccc}\hline\hline
%   Scenario  & MSE & OMSE  & TP & FP & 95\% Cov. Prob. 1&95\% Cov. Prob. 2      \\
%\hline
%    $\rho=0,\sigma=0.1$ &0.73  & 0.64 & 2&0.15&0.80&0.83\\
%    $\rho=0.2,\sigma=0.1$ &0.70 & 0.63 &2 &0.16 &0.75&0.84\\
%    $\rho=0.5,\sigma=0.1$ &0.97 & 0.85  &2 &0.15 &0.78&0.84\\
%    $\rho=0,\sigma=0.3$ & 1.73 & 1.47 & 2 &0.21&0.73&0.78 \\
%    $\rho=0.2,\sigma=0.3$ &2.06 & 1.81 & 2 & 0.16 &0.60 &0.75\\
%    $\rho=0.5,\sigma=0.3$ &2.75 & 2.26 & 2 &0.26 & 0.67&0.78\\
%   \hline
%  \end{tabular}}
%%      \label{symbols}
%\end{table}

\begin{table}
  \caption{Simulation results for penalized multiple functional regression. \label{tab:sim}}
\bigskip
\centering{\begin{tabular}{ccccccc}\hline\hline
   Scenario  & MSE & OMSE  & TP & FP & 95\% Cov.Prob.1&95\% Cov.Prob.2      \\
\hline
    $\rho=0.0,\sigma=0.1$ &0.73  & 0.64 & 2&0.08&0.92&0.93\\
    $\rho=0.2,\sigma=0.1$ &0.70 & 0.63 &2 &0.09 &0.92&0.94\\
    $\rho=0.5,\sigma=0.1$ &0.97 & 0.85  &2 &0.08 &0.93&0.94\\
    $\rho=0.0,\sigma=0.3$ & 1.73 & 1.47 & 1.77 &0.13&0.93&0.94 \\
    $\rho=0.2,\sigma=0.3$ &2.06 & 1.81 & 1.81 & 0.11 &0.93 &0.94\\
    $\rho=0.5,\sigma=0.3$ &2.75 & 2.26 & 1.80 &0.15 & 0.92&0.92\\
   \hline
  \end{tabular}}
%      \label{symbols}
\end{table}

\subsection{Spectrometrics Data}
\begin{figure}
\centerline{\subfigure[]{\includegraphics[width=2.5in]{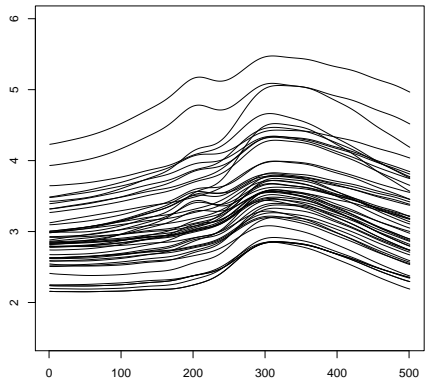}
}
}
\caption{The spectrometric curves. \label{fig:spec}   }
\end{figure}

We illustrate our approach on the real spectrometrics dataset, which contains 215
spectra of light absorbance for meat samples as functions of the wavelengths.
Because of the denseness of wavelengths at which the measurements are made,
the subjects are naturally treated as continuous curves. Figure \ref{fig:spec} shows the first 50 curves in the dataset. This dataset has been
previously used in functional nonparametric regression studies where the covariate is the
spectra curve and the response is the percentage of fat content in the piece of
meat \citep{ferraty02,ferraty06,ferraty07}. In nonparametric kernel regression, as shown in the above mentioned works, choice of semi-metric which defined the notion of distance between curves is crucial for the performance of the estimator. Previous study suggested that for nonparametric regression function estimation, taking as the semi-metric the $L_2$ distance between the
second derivatives of the spectra gives favorable results based on its performance on hold-out validation data. A desirable feature of an estimation procedure would be to determine the appropriate order of derivative automatically.

\begin{figure}
\centerline{\subfigure[]{\includegraphics[width=2.5in]{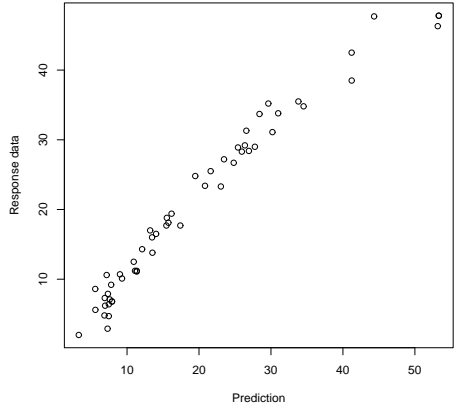}
}
}
\caption{Prediction accuracy with penalized multiple functional regression on 55 validation samples.\label{fig:predict}   }
\end{figure}

\begin{figure}
\centerline{\subfigure[]{\includegraphics[width=2.5in]{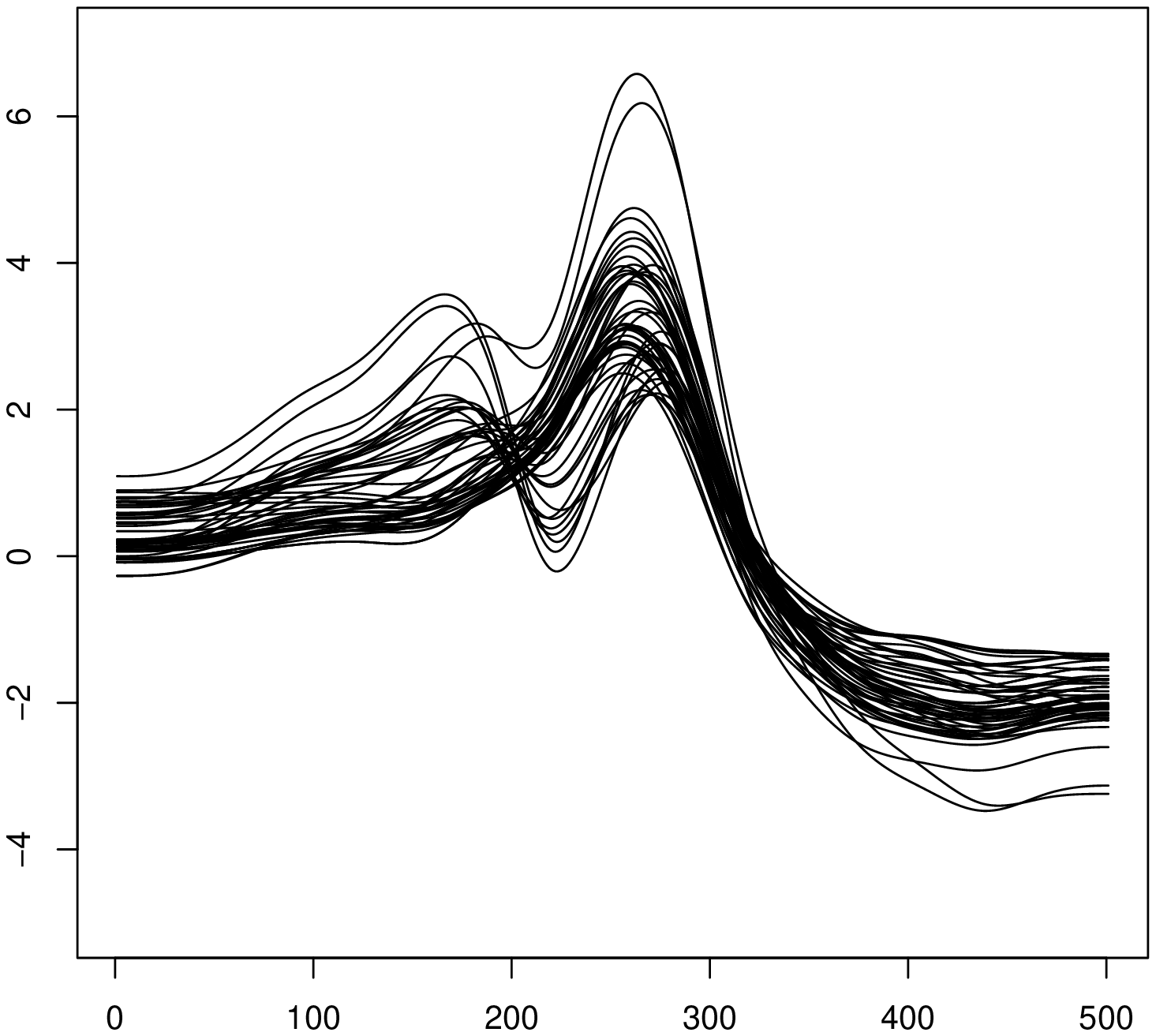}
}
\hfil
\subfigure[]{\includegraphics[width=2.5in]{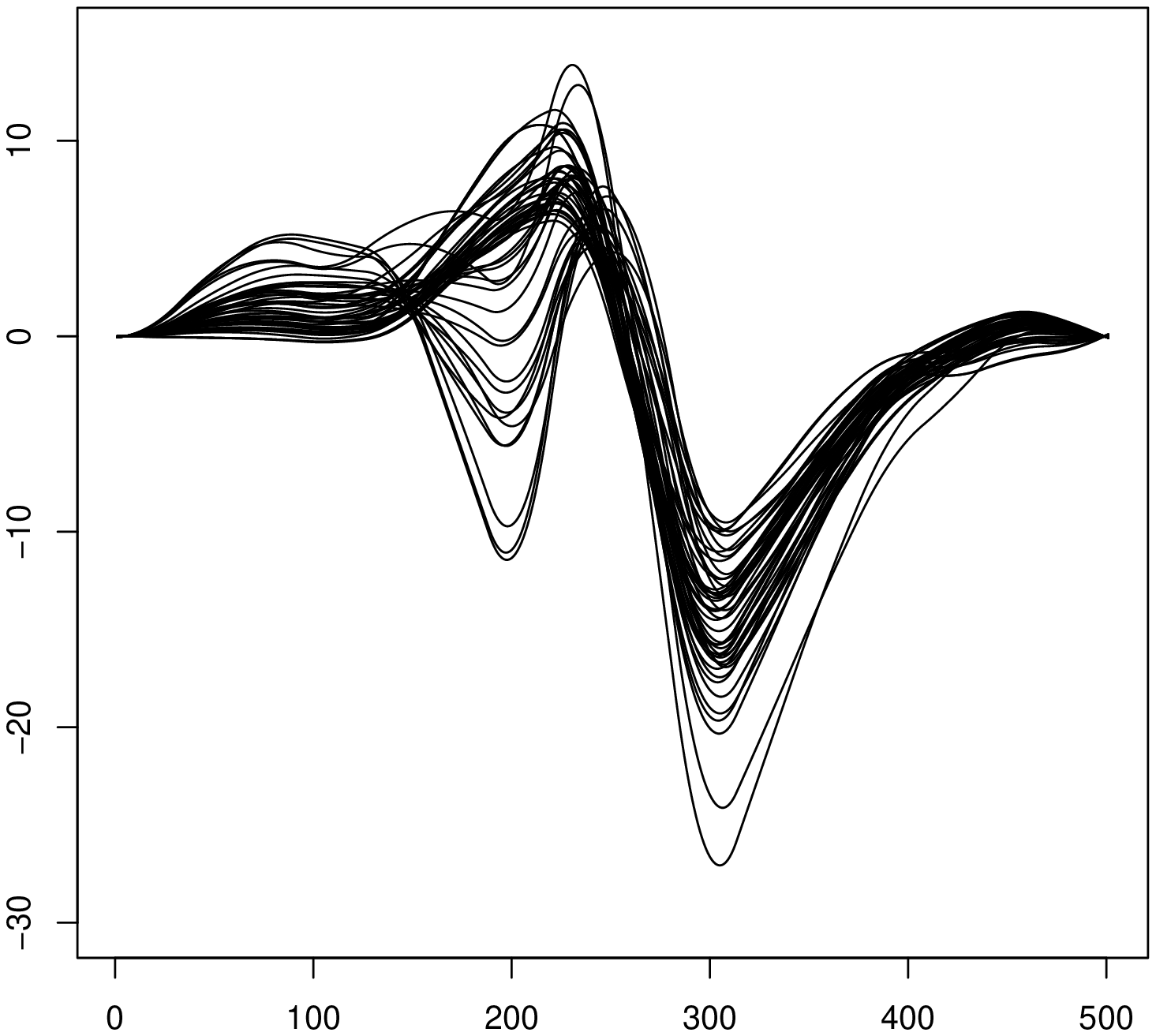}
}}
\centerline{\subfigure[]{\includegraphics[width=2.5in]{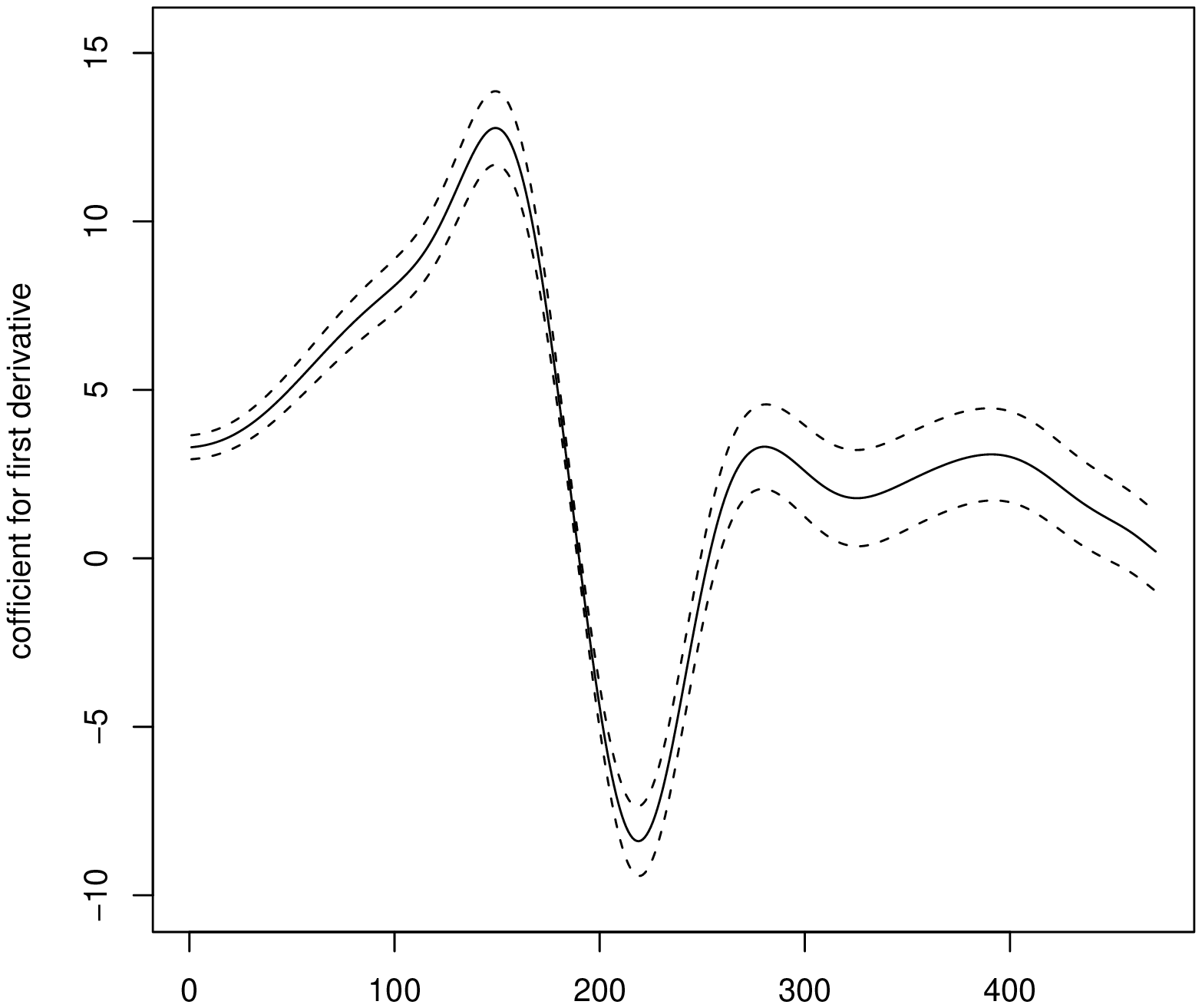}
}
\hfil
\subfigure[]{\includegraphics[width=2.5in]{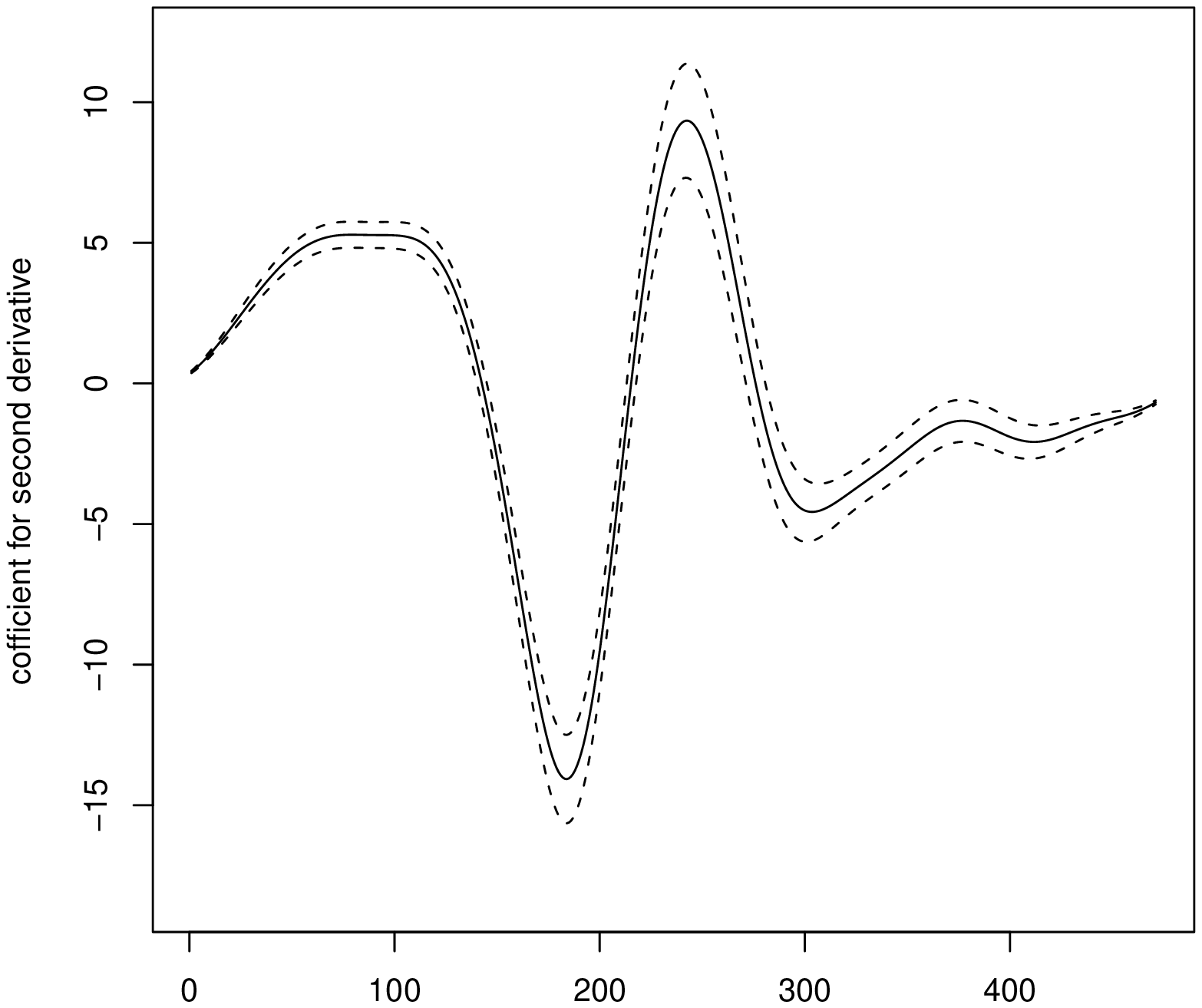}
}}
\caption{(a) and (b): 1st and 2nd derivative of the spectrometric data. Only 50 samples are shown in the figure. (c) and (d): Estimated functional linear coefficient corresponding to 1st and 2nd derivative curves respectively, with 95\% pointwise confidence interval shown as dotted lines. \label{fig:estimate}   }
\end{figure}

Here we apply the multiple functional linear regression model to the spectrometrics data. We treat the original function itself as well as up to its 3rd derivatives as the predictors in our model. The idea of using different orders of derivatives of curves as covariates in the functional linear model is similar to using transformations of the original covariates in classical multiple linear regression, which makes the linear model more flexible.   Compared to nonparametric functional kernel regression, the functional linear model is more easily interpretable and thus an interesting alternative. For this data, we train on the first 160 spectra and use the rest as validation. We examine the prediction accuracy of the estimated model using mean squared error on the validation data, defined as
\[MSE=\frac{1}{55}\sum_{i=161}^{215} (Y_i-\hat{Y}_i)^2.\]
With the smoothing parameters selected by GCV, the relevant predictors are found to be the 1st and 2nd derivatives of the spectra curves, achieving an MSE of $8.31$. Figure \ref{fig:predict} clearly shows the ability of the estimated model to predict the responses. For comparison, we also computed the nonparametric kernel regression using the \textbf{funopare.kernel.cv} function provided in the \textbf{npfda} package (it uses cross-validation to select the bandwidth), which gives a smaller MSE of $5.37$.  However, when using functional linear modeling, unlike kernel regression, we can visually examine the features of the functional coefficients for interpretation. For example, from Figure \ref{fig:estimate}, higher fat content is seen to be related to higher values around point $160$ and lower values around point $215$ in the first derivative, as well as lower values around point $190$ in the second derivative.

\section{Concluding Remarks}
We propose in this article a regularization method for shrinkage estimation of multiple functional linear regression models. We have shown that the proposed method is consistent in estimation and variable selection. A computational algorithm based on local quadratic approximation is proposed. It is also possible to use local linear approximation \citep{zou08} and our choice made here is based on ease of implementation since closed form solution exists for each iteration. Our simulation results demonstrated the effectiveness of the method and the application to spectrometrics data provides an interesting alternative perspective to the previously used kernel regression on this data. 

We would like to finish this paper by discussing some possible topics for future study. One possibility is to consider partially functional linear regression where scalar covariates are considered simultaneously. Variable selection can be applied to both the functional and non-functional part. Another direction is to consider multiple functional linear regression when the number of predictors diverges with sample size. How to extend the shrinkage estimation results to generalized functional linear model \citep{james02,muller05,cardot05} is another interesting topic for further study.

\section*{Appendix}
The following two lemmas study some properties of the estimated principal component scores and are important for the proof of Theorem \ref{thm:main}. Throughout the appendix, we follow the notations and assumptions in the main text.
\begin{lemma}\label{lem:1}
We have $|\hat{\xi}_{ijk}-\xi_{ijk}|=O_p(K^{\alpha+1}/\sqrt{n})$ and $|\sum_{i=1}^n\hat{\xi}_{ij_1k_1}\hat{\xi}_{ij_2k_2}/n-\lambda^{j_1,j_2}_{k_1,k_2}|=O_p(K^{\alpha+1}/\sqrt{n})$.
\end{lemma}
\textit{Proof.} Since $\xi_{ijk}=\int (X_{ij}-\mu_j){\phi}_{jk}$ and $\hat{\xi}_{ijk}=\int (X_{ij}-\bar{X}_j)\hat{\phi}_{jk}$, we have $|\hat{\xi}_{ijk}-\xi_{ijk}|^2=O_p(||\bar{X}_j-\mu_j||^2+||\hat{\phi}_{jk}-\phi_{jk}||^2)=O_p(K^{2\alpha+2}/n)$, using assumption (c2) and equation (5.2) in \cite{hall07}. For the second part, we have 
\begin{eqnarray*}
&&\frac{\sum_{i=1}^n\hat{\xi}_{ij_1k_1}\hat{\xi}_{ij_2k_2}}{n}-\lambda^{j_1,j_2}_{k_1,k_2}\\
&=&\left(\frac{\sum_{i=1}^n\hat{\xi}_{ij_1k_1}\hat{\xi}_{ij_2k_2}}{n}-\frac{\sum_{i=1}^n{\xi}_{ij_1k_1}{\xi}_{ij_2k_2}}{n}\right)+\left(\frac{\sum_{i=1}^n{\xi}_{ij_1k_1}{\xi}_{ij_2k_2}}{n}-\lambda^{j_1,j_2}_{k_1,k_2}\right)\\
&=:& (I)+(II).
\end{eqnarray*}
Obviously the second term is of order $O_p(n^{-1/2})$. The first term above is further decomposed as
\begin{eqnarray*}
(I)&=& \frac{1}{n}\sum_{i=1}^n\left[(\hat{\xi}_{ij_1k_1}-\xi_{ij_1k_1})\hat{\xi}_{ij_2k_2}+(\hat{\xi}_{ij_2k_2}-\xi_{ij_2k_2}){\xi}_{ij_1k_1}\right].\\
\end{eqnarray*}
Using $|\hat{\xi}_{ijk}|=O_p(1)$ since $|\hat{\xi}_{ij_1k_1}-\xi_{ij_1k_1}|=o_p(1)$, we have a bound $O_p(K^{\alpha+1}/\sqrt{n})$ for (I), and the proof is completed. $\Box$
\begin{lemma} \label{lem:2} For any $A$ denoting a subset of $\{1,2,\ldots,p\}$, let $\hat{Z}_A$ be the columns of $\hat{Z}$ corresponding to those predictors in $A$, and similarly let $\Lambda_A$ be the submatrix of $\Lambda$ corresponding to the predictors in $A$, then the minimum eigenvalue of $\hat{Z}_A^T\hat{Z}_A/n$ is lower bounded by a constant multiple of $K^{-\alpha}$, i.e. $\rho_{\min}(\hat{Z}_A^T\hat{Z}_A/n)=\Omega_p(K^{-\alpha})$.
\end{lemma}
\textit{Proof.} We will use $||\cdot||$ to denote also the operator norm of a matrix, and use $||\cdot||_1$ for maximum row sum of a matrix. Since $|\rho_{\min}(\hat{Z}_A^T\hat{Z}_A/n)-\rho_{\min}(\Lambda_A)|\le ||\hat{Z}_A^T\hat{Z}_A/n-\Lambda_A||\le ||\hat{Z}_A^T\hat{Z}_A/n-\Lambda_A||_1=O_p(K^{\alpha+2}/\sqrt{n})$ using the previous lemma. This together with assumption (c5) implies the statement of the lemma. $\Box$\\

\textbf{Proof of Theorem \ref{thm:main}.} In the proof we denote the minimum eigenvalue of $\hat{Z}^T\hat{Z}/n$ by $\rho^*$ and thus $\rho^*=\Omega_p(K^{-\alpha})$ by Lemma \ref{lem:2}. The true functional coefficients are denoted by $\beta_j=\sum_{k}b_{jk}\phi_{jk}$. Then
\begin{eqnarray*}
0&\ge&J(\hat{b})-J(b)\\
&=&||\mathbf{Y}-\bar{Y}\mathbf{1}-\hat{Z}\hat{b}||^2-||\mathbf{Y}-\bar{Y}\mathbf{1}-\hat{Z}b||^2+n\sum_jp_\lambda(||\hat{b}_j||)-n\sum_jp_\lambda(||b_j||)\\
&=&||\mathbf{Y}-\bar{Y}\mathbf{1}-\hat{Z}{b}+\hat{Z}{b}-\hat{Z}\hat{b}||^2-||\mathbf{Y}-\bar{Y}\mathbf{1}-\hat{Z}b||^2+n\sum_jp_\lambda(||\hat{b}_j||)-n\sum_jp_\lambda(||b_j||)\\
&=&2(\mathbf{Y}-\bar{Y}\mathbf{1}-\hat{Z}b)^T\hat{Z}(b-\hat{b})+||\hat{Z}(b-\hat{b})||^2+n\sum_jp_\lambda(||\hat{b}_j||)-n\sum_jp_\lambda(||b_j||).
\end{eqnarray*}
Let $\eta=\hat{Z}(\hat{Z}^T\hat{Z})^{-1}\hat{Z}^T(\mathbf{Y}-\bar{Y}\mathbf{1}-\hat{Z}b)$ be the projection of $\mathbf{Y}-\bar{Y}\mathbf{1}-\hat{Z}b$ onto the columns of $\hat{Z}$, Lemma \ref{lem:3} below shows that $||\eta||^2=O_p(r_n^2)$, where $r_n^2=O_p(K^{2\alpha+3}+nK^{-\alpha-2\beta+1})$. By assumption (c3), we get $nK^{-\alpha-2\beta+1}=O(K^{2\alpha+3})$ and thus $r_n^2=O(K^{2\alpha+3})$ and the above displayed equation can be continued as
\begin{eqnarray}
0&\ge& -O_p(r_n)||\hat{Z}(b-\hat{b})||+||\hat{Z}(b-\hat{b})||^2+n\sum_jp_\lambda(||\hat{b}_j||)-n\sum_jp_\lambda(||b_j||)\nonumber\\
&\ge&-O_p(r_n^2)-\frac{1}{2}||\hat{Z}(b-\hat{b})||^2+||\hat{Z}(b-\hat{b})||^2+n\sum_jp_\lambda(||\hat{b}_j||)-n\sum_jp_\lambda(||b_j||)\label{eqn:rate}\\
&\ge&-O_p(r_n^2)+n\rho^*||b-\hat{b}||^2-n\lambda\sum_j||\hat{b}_j-b_j||\nonumber\\
&\ge&-O_p(r_n^2)+n\rho^*||b-\hat{b}||^2-\frac{2n\lambda^2}{\rho^*}-\frac{n\rho^*}{2}||\hat{b}-b||^2,\nonumber
\end{eqnarray}
where we used Cauchy-Schwartz inequality on the second line, the property $|p_\lambda(a)-p_\lambda(b)|\le\lambda|a-b|$ on the third line, and Cauchy-Schwartz inequality again on the last line. Thus $||\hat{b}-b||^2=O_p(\frac{r_n^2}{n\rho^*}+\frac{\lambda^2}{(\rho^*)^2})=o_p(1)$ by assumptions (c3) and (c4).

The convergence rate for $||\hat{b}-b||^2$ can be improved to $O_p(r_n^2/n\rho^*)$, which is useful in the proof of part (b). Since $||\hat{b}-b||=o_p(1)$ and $\lambda\rightarrow 0$, we have $P(p_\lambda(||\hat{b}_j||)=p_\lambda(||{b}_j||),1\le j\le s)\rightarrow 1$ and thus $\sum_jp_\lambda(||\hat{b}_j||)-\sum_jp_\lambda(||{b}_j||)\ge 0$ with probability converging to 1. This combined with (\ref{eqn:rate}) gives $||\hat{b}-b||^2=O_p(r_n^2/n\rho^*)$.

From $||\hat{b}_j-b_j||=o_p(1)$, part (a) is easily shown using the following decomposition
\begin{eqnarray*}
||\hat{\beta}_j-\beta_j||^2&\le& 3||\hat{b}_j-b_j||^2+3\int\left[\sum_{k=1}^Kb_{jk}(\hat{\phi}_{jk}-\phi_{jk})\right]^2+3\sum_{k=K+1}^\infty b_{jk}^2\\
&=&3||\hat{b}_j-b_j||^2+3K\sum_{k=1}^Kb_{jk}^2||\hat{\phi}_{jk}-\phi_{jk}||^2+3\sum_{k=K+1}^\infty b_{jk}^2\\
&=&3||\hat{b}_j-b_j||^2+O(K\cdot\frac{K^{2\alpha+2}}{n})+3\sum_{k=K+1}^\infty b_{jk}^2,
\end{eqnarray*}
and obviously all the terms in the above display converge to zero.

Now we prove part (b) of the theorem. Let $\hat{b}^*=(\hat{b}_1^T,\ldots,\hat{b}_s^T,0,\ldots,0)^T$, that is, $\hat{b}^*$ is obtained from $\hat{b}$ by constraining the truly irrelevant components to be zero. %Suppose there exists $j_0>s$ such that $\hat{b}_j\neq 0$. 
By similar arguments for the proof of part (a), we have
\begin{eqnarray}
0&\ge&J(\hat{b})-J(\hat{b}^*)\nonumber\\
&=&2(\mathbf{Y}-\bar{Y}\mathbf{1}-\hat{Z}\hat{b}^*)^T\hat{Z}(\hat{b}-\hat{b}^*)+||\hat{Z}(\hat{b}-\hat{b}^*)||^2+n\sum_jp_\lambda(||\hat{b}_j||)-n\sum_jp_\lambda(||\hat{b}^*_j||)\nonumber\\
&\ge&-O_p(||\eta^*||)||\hat{Z}(\hat{b}-\hat{b}^*)||+n\sum_{j=s+1}^p p_\lambda(||\hat{b}_j||)\nonumber\\
&\ge&-O_p(||\eta^*||)\sqrt{n}\sum_{j=s+1}^p||\hat{b}_j||+n\lambda\sum_{j=s+1}^p||\hat{b}_j||,\label{eqn:partb}
\end{eqnarray}
where $\eta^*=\hat{Z}(\hat{Z}^T\hat{Z})^{-1}\hat{Z}^T(\mathbf{Y}-\bar{Y}\mathbf{1}-\hat{Z}\hat{b}^*)$. In the last line above we use the fact that $||\hat{b}||_j=O_p(r_n/\sqrt{n\rho^*})=o_p(\lambda)$ when $j>s$ (from the proof of part (a)) and thus $p_\lambda(||\hat{b}_j||)=\lambda||\hat{b}_j||$.

We bound $||\eta^*||$ as follows. 
\begin{eqnarray*}
||\eta^*||^2&\le&2||\eta||^2+2||\hat{Z}(\hat{b}^*-b)||^2\\
&=&O_p(r_n^2)+O_p(nr_n^2/(n\rho^*))=O_p(r_n^2/\rho^*).
\end{eqnarray*}
Since we have that $O_p(||\eta^*||)=o_p(\sqrt{n}\lambda)$, we will have a contradiction in (\ref{eqn:partb}) if $\sum_{j=s+1}^p||\hat{b}_j||>0$. $\Box$\\

\begin{lemma}\label{lem:3}
Let $\eta=\hat{Z}(\hat{Z}^T\hat{Z})^{-1}\hat{Z}^T(\mathbf{Y}-\bar{Y}\mathbf{1}-\hat{Z}b)$ as in the proof of Theorem \ref{thm:main}, then $||\eta||^2=O_p(r_n^2)$ where $r_n^2=K^{2\alpha+3}+nK^{-\alpha-2\beta+1}$.
\end{lemma}
\textit{Proof.} Denote by $Z$ the matrix similar in structure to $\hat{Z}$ but contains the true principal component scores $\xi_{ijk}$ instead of $\hat{\xi}_{ijk}$. We have the decomposition
\begin{equation}\label{eqn:etadecomp}
\mathbf{Y}-\bar{Y}\mathbf{1}-\hat{Z}b=\epsilon+(\mu-\bar{Y})\mathbf{1}+(Z-\hat{Z})b+\nu,
\end{equation}
where $\epsilon=(\epsilon_1,\ldots,\epsilon_n)^T$ and $\nu$ is a $n$-dimensional vector with $i$-th component given by
\[\nu_i=\sum_{j=1}^p\sum_{k=K+1}^\infty\xi_{ijk}b_{jk}.\]
Let $P_{\hat{Z}}=\hat{Z}(\hat{Z}^T\hat{Z})^{-1}\hat{Z}^T$. Now $\eta=P_{\hat{Z}}(\mathbf{Y}-\bar{Y}\mathbf{1}-\hat{Z}b)$ is the projection of the four terms in the decomposition (\ref{eqn:etadecomp}) onto columns of $\hat{Z}$, and we bound each term in turn below.

Since $||P_{\hat{Z}}\epsilon||^2=\epsilon^TP_{\hat{Z}}\epsilon$,
using the fact $E[\epsilon^TP_{\hat{Z}}\epsilon|X]=\sigma^2tr(P_{\hat{Z}})=\sigma^2pK$, $Var(\epsilon^TP_{\hat{Z}}\epsilon|X)=2\sigma^4tr(P_{\hat{Z}}^2)+(E\epsilon_i^4-3\sigma^2)\sum_{j=1}^{n}(P_{\hat{Z}})_{jj}^2\le 2\sigma^4tr(P_{\hat{Z}}^2)+|E\epsilon_i^4-3\sigma^2|\sum_{j=1}^{n}(P_{\hat{Z}})_{jj}=O_p(K)$ (see for example equations (3.3), (3.4) in \cite{huangfan99}), where $(P_{\hat{Z}})_{jj}$ are the diagonal elements of $P_{\hat{Z}}$ which are all no larger than $1$ since $P_{\hat{Z}}$ is a projection matrix. Using the equalities $E\epsilon^TP_{\hat{Z}}\epsilon=E[E(\epsilon^TP_{\hat{Z}}\epsilon|X)]$ and $Var(\epsilon^TP_{\hat{Z}}\epsilon)=E[Var(\epsilon^TP_{\hat{Z}}\epsilon|X)]+Var(E[\epsilon^TP_{\hat{Z}}\epsilon|X])$, we have 
\begin{equation}\label{eqn:term1}
||P_{\hat{Z}}\epsilon||^2=O_p(K).
\end{equation}

Besides,
\begin{eqnarray*}
||P_{\hat{Z}}(Z-\hat{Z})b||^2&\le&||(Z-\hat{Z})b||^2=O(||(Z-\hat{Z})^T(Z-\hat{Z})||).
\end{eqnarray*}
Using Lemma \ref{lem:1}, we get $||(Z-\hat{Z})^T(Z-\hat{Z})||\le ||(Z-\hat{Z})^T(Z-\hat{Z})||_1=O_p(K^{2\alpha+3})$ and thus 
\begin{equation}\label{eqn:term2}
||P_{\hat{Z}}(Z-\hat{Z})b||^2=O_p(K^{2\alpha+3}).
\end{equation}

Finally, 
\begin{eqnarray*}
Var(\sum_{k=K+1}^\infty\xi_{ijk}b_{jk})&=&\sum_{k=K+1}^\infty\lambda_{jk}b_{jk}^2\\
&=&O(\sum_{k=K+1}^\infty k^{-\alpha}k^{-2\beta})\\
&=&O(K^{-\alpha-2\beta+1})
\end{eqnarray*}
Since the number of predictors $p$ is fixed, we have $Var(\nu_i)=O(K^{-\alpha-2\beta+1})$ and thus 
\begin{equation}\label{eqn:term3}
||\nu||^2=O_p(nK^{-\alpha-2\beta+1}).
\end{equation}
Combining (\ref{eqn:term1}), (\ref{eqn:term2}), (\ref{eqn:term3}) as well as $|\mu-\bar{Y}|=O_p(n^{-1/2})$, we get $||\eta||^2=O_p(r_n^2)$. $\Box$

\textbf{Proof of Proposition \ref{prop:1}.} From equation (\ref{eqn:trans}), the Karhunen-Lo\`{e}ve expansion of the predictors are given by 
\[X_i(t)=\sum_{k=1}^\infty\xi_{ik}\phi_k(t), \mbox { with } \xi_{ik}=\sum_{j=1}^la_{ij}\omega_{jk}, i=1,2.\]
Using the notation in the main text, we have that the general entries of $\Lambda$ are given by 
\[\lambda^{i_1,i_2}_{k_1,k_2}=E\xi_{i_1k_1}\xi_{i_2k_2}
	=\left\{\begin{array}{cc}
		\sum_{j=1}^la_{i_1j}a_{i_2j}\kappa_{jk}& k_1=k_2=k\\
		0 & k_1\neq k_2\;.\\
		\end{array}\right.
\]
Thus in this case, in the block matrix form,
\[\Lambda=\left(\begin{array}{cc}
		E&F\\
		F^T&G
		\end{array}\right),\]
and the matrix $F$ is also diagonal. Since $\Lambda$ is similar to the matrix
\[\tilde{\Lambda}=\left(\begin{array}{cc}
		E&0\\
		0&G-F^TE^{-1}F
		\end{array}\right),\]
the eigenvalues of $\Lambda$ are just the diagonal elements of $E$ and $G-F^TE^{-1}F$. The eigenvalues of $E$ are $\Omega(K^{-\alpha})$ by assumption, and the diagonal elements of $G-F^TE^{-1}F$ are 
\begin{eqnarray*}
&&\sum_{j=1}^la_{2j}^2\kappa_{jk}-\frac{(\sum_{j=1}^la_{1j}a_{2j}\kappa_{jk})^2}{\sum_{j=1}^la_{1j}^2\kappa_{jk}}\\
&=&\frac{\sum_{1\le j_1\neq _2\le l}(a_{1j_1}a_{2j_2}\sqrt{\kappa_{j_1k}\kappa_{j_2k}}-a_{2j_1}a_{1j_2}\sqrt{\kappa_{j_1k}\kappa_{j_2k}})^2  }{2\sum_{j=1}^la_{1j}^2\kappa_{jk}}\\
&\ge&\frac{c^2\sum_{1\le j_1\neq j_2\le l}(a_{1j_1}a_{2j_2}-a_{2j_1}a_{1j_2})^2k^{-2\alpha}    }{2C\sum_{j=1}^la_{1j}^2k^{-\alpha}}\\
&=&\Omega(k^{-\alpha})
\end{eqnarray*}
$\Box$

\bibliographystyle{asa}
\bibliography{papers.txt,books.txt}

\begin{thebibliography}{35}
\newcommand{\enquote}[1]{``#1''}
\expandafter\ifx\csname natexlab\endcsname\relax\def\natexlab#1{#1}\fi

\bibitem[{Cai and Hall(2006)}]{cai06}
Cai, T.~T. and Hall, P. (2006), \enquote{Prediction in functional linear
  regression,} \textit{Annals of Statistics}, 34, 2159--2179.

\bibitem[{Cardot et~al.(1999)Cardot, Ferraty, and Sarda}]{cardot99}
Cardot, H., Ferraty, F., and Sarda, P. (1999), \enquote{Functional linear
  model,} \textit{Statistics \& Probability Letters}, 45, 11--22.

\bibitem[{Cardot and Sarda(2005)}]{cardot05}
Cardot, H. and Sarda, P. (2005), \enquote{Estimation in generalized linear
  models for functional data via penalized likelihood,} \textit{Journal of
  Multivariate Analysis}, 92, 24--41.

\bibitem[{Carroll et~al.(2009)Carroll, Delaigle, and Hall}]{carroll09}
Carroll, R.~J., Delaigle, A., and Hall, P. (2009), \enquote{Nonparametric
  prediction in measurement error models,} \textit{Journal of the American
  Statistical Association}, 104, 993--1003.

\bibitem[{Crambes et~al.(2009)Crambes, Kneip, and Sarda}]{crambes09}
Crambes, C., Kneip, A., and Sarda, P. (2009), \enquote{Smoothing splines
  estimators for functional linear regression,} \textit{Annals of Statistics},
  37, 35--72.

\bibitem[{Fan and Li(2001)}]{fan01}
Fan, J.~Q. and Li, R.~Z. (2001), \enquote{Variable selection via nonconcave
  penalized likelihood and its oracle properties,} \textit{Journal of the
  American Statistical Association}, 96, 1348--1360.

\bibitem[{Fan and Peng(2004)}]{fan04}
Fan, J.~Q. and Peng, H. (2004), \enquote{Nonconcave penalized likelihood with a
  diverging number of parameters,} \textit{Annals of Statistics}, 32, 928--961.

\bibitem[{Ferraty et~al.(2007)Ferraty, Mas, and Vieu}]{ferraty07}
Ferraty, F., Mas, A., and Vieu, P. (2007), \enquote{Nonparametric regression on
  functional data: Inference and practical aspects,} \textit{Australian \& New
  Zealand Journal of Statistics}, 49, 267--286.

\bibitem[{Ferraty and Vieu(2002)}]{ferraty02}
Ferraty, F. and Vieu, P. (2002), \enquote{The functional nonparametric model
  and application to spectrometric data,} \textit{Computational Statistics},
  17, 545--564.

\bibitem[{Ferraty and Vieu(2006)}]{ferraty06}
--- (2006), \textit{Nonparametric functional data analysis: theory and
  practice}, Springer series in statistics, New York, NY: Springer.

\bibitem[{Hall and Horowitz(2007)}]{hall07}
Hall, P. and Horowitz, J.~L. (2007), \enquote{Methodology and convergence rates
  for functional linear regression,} \textit{Annals of Statistics}, 35, 70--91.

\bibitem[{Huang et~al.(2008)Huang, Horowitz, and Ma}]{huangma08}
Huang, J., Horowitz, J.~L., and Ma, S.~G. (2008), \enquote{Asymptotic
  properties of bridge estimators in sparse high-dimensional regression
  models,} \textit{Annals of Statistics}, 36, 587--613.

\bibitem[{Huang and Fan(1999)}]{huangfan99}
Huang, L.~S. and Fan, J.~Q. (1999), \enquote{Nonparametric estimation of
  quadratic regression functionals,} \textit{Bernoulli}, 5, 927--949.

\bibitem[{James(2002)}]{james02}
James, G.~M. (2002), \enquote{Generalized linear models with functional
  predictors,} \textit{Journal of the Royal Statistical Society Series
  B-Statistical Methodology}, 64, 411--432.

\bibitem[{Lian(2007)}]{lian07}
Lian, H. (2007), \enquote{Nonlinear functional models for functional responses
  in reproducing kernel Hilbert spaces,} \textit{Canadian Journal of
  Statistics-Revue Canadienne De Statistique}, 35, 597--606.

\bibitem[{Liang and Li(2009)}]{liang09}
Liang, H. and Li, R.~Z. (2009), \enquote{Variable selection for partially
  linear models with measurement errors,} \textit{Journal of the American
  Statistical Association}, 104, 234--248.

\bibitem[{Lin and Zhang(2006)}]{lin06}
Lin, Y. and Zhang, H.~H. (2006), \enquote{Component selection and smoothing in
  multivariate nonparametric regression,} \textit{Annals of Statistics}, 34,
  2272--2297.

\bibitem[{Meinshausen and Buhlmann(2006)}]{meinshausen06}
Meinshausen, N. and Buhlmann, P. (2006), \enquote{High-dimensional graphs and
  variable selection with the Lasso,} \textit{Annals of Statistics}, 34,
  1436--1462.

\bibitem[{Muller and Stadtmuller(2005)}]{muller05}
Muller, H.~G. and Stadtmuller, U. (2005), \enquote{Generalized functional
  linear models,} \textit{Annals of Statistics}, 33, 774--805.

\bibitem[{Preda(2007)}]{preda07}
Preda, C. (2007), \enquote{Regression models for functional data by reproducing
  kernel Hilbert spaces methods,} \textit{Journal of Statistical Planning and
  Inference}, 137, 829--840.

\bibitem[{Ramsay and Silverman(2005)}]{ramsay05}
Ramsay, J.~O. and Silverman, B.~W. (2005), \textit{Functional data analysis},
  Springer series in statistics, New York: Springer, 2nd ed.

\bibitem[{Tibshirani(1996)}]{tibshirani96}
Tibshirani, R. (1996), \enquote{Regression shrinkage and selection via the
  Lasso,} \textit{Journal of the Royal Statistical Society Series
  B-Methodological}, 58, 267--288.

\bibitem[{Wahba(1990)}]{wahba90}
Wahba, G. (1990), \textit{Spline models for observational data}, Philadelphia,
  PA: Society for Industrial and Applied Mathematics.

\bibitem[{Wang and Leng(2007)}]{wangleng07}
Wang, H.~S. and Leng, C.~L. (2007), \enquote{Unified LASSO estimation by least
  squares approximation,} \textit{Journal of the American Statistical
  Association}, 102, 1039--1048.

\bibitem[{Wang and Xia(2009)}]{wang09}
Wang, H.~S. and Xia, Y.~C. (2009), \enquote{Shrinkage estimation of the varying
  coefficient model,} \textit{Journal of the American Statistical Association},
  104, 747--757.

\bibitem[{Wang et~al.(2007)Wang, Chen, and Li}]{wang07}
Wang, L.~F., Chen, G., and Li, H.~Z. (2007), \enquote{Group SCAD regression
  analysis for microarray time course gene expression data,}
  \textit{Bioinformatics}, 23, 1486--1494.

\bibitem[{Wang et~al.(2008)Wang, Li, and Huang}]{wang08}
Wang, L.~F., Li, H.~Z., and Huang, J. H.~Z. (2008), \enquote{Variable selection
  in nonparametric varying-coefficient models for analysis of repeated
  measurements,} \textit{Journal of the American Statistical Association}, 103,
  1556--1569.

\bibitem[{Yao et~al.(2005)Yao, Muller, and Wang}]{yao05}
Yao, F., Muller, H.~G., and Wang, J.~L. (2005), \enquote{Functional linear
  regression analysis for longitudinal data,} \textit{Annals of Statistics},
  33, 2873--2903.

\bibitem[{Yuan and Cai(2010+)}]{yuan09}
Yuan, M. and Cai, T.~T. (2010+), \enquote{A reproducing kernel Hilbert space
  approach to functional linear regression,} \textit{Annals of Statistics}, to
  appear.

\bibitem[{Yuan and Lin(2006)}]{yuan06}
Yuan, M. and Lin, Y. (2006), \enquote{Model selection and estimation in
  regression with grouped variables,} \textit{Journal of the Royal Statistical
  Society Series B-Statistical Methodology}, 68, 49--67.

\bibitem[{Zhang(2006)}]{zhang06}
Zhang, H.~H. (2006), \enquote{Variable selection for support vector machines
  via smoothing spline ANOVA,} \textit{Statistica Sinica}, 16, 659--674.

\bibitem[{Zhao and Yu(2006)}]{zhaoyu06}
Zhao, P. and Yu, B. (2006), \enquote{On model selection consistency of Lasso,}
  \textit{Journal of Machine Learning Research}, 7, 2541--2563.

\bibitem[{Zhu et~al.(2010+)Zhu, Vannucci, and Cox}]{zhu10}
Zhu, H., Vannucci, M., and Cox, D. (2010+), \enquote{A Bayesian Hierarchical
  Model for Classification with Selection of Functional Predictors,}
  \textit{Biometrics}, to appear.

\bibitem[{Zou(2006)}]{zou06}
Zou, H. (2006), \enquote{The adaptive lasso and its oracle properties,}
  \textit{Journal of the American Statistical Association}, 101, 1418--1429.

\bibitem[{Zou and Li(2008)}]{zou08}
Zou, H. and Li, R.~Z. (2008), \enquote{One-step sparse estimates in nonconcave
  penalized likelihood models,} \textit{Annals of Statistics}, 36, 1509--1533.

\end{thebibliography}

\end{document}